\documentclass[prl,twocolumn,nofootinbib, preprintnumbers, superscriptaddress]{revtex4-1}

\usepackage{amsmath,amssymb,amscd,simplewick}
\usepackage{listings}
\usepackage{dsfont}
\usepackage{slashed}
\usepackage{color}
\usepackage{ulem}

\usepackage{graphicx}
\usepackage{epstopdf}
\usepackage{subfigure}
\usepackage{epsfig}

\usepackage{xcolor}
\usepackage[colorlinks=true,
            linkcolor=blue,
            urlcolor=blue,
            citecolor=green,          
            bookmarks=true,
            bookmarksnumbered=true,
            breaklinks=true,
            pdfpagemode=Fullscreen,
            pdfstartview=FitBH]{hyperref}

\hypersetup{pdfauthor = {Shao-Feng Ge},
	     pdftitle = {}, 
	     pdfsubject = {}, 
             pdfkeywords = {}, 
	     pdfcreator = {LaTeX with hyperref package},
	     pdfproducer = {dvips + ps2pdf} }

\definecolor{gesfpurple}{rgb}{0.47,0.19,0.42}

\definecolor{gesflanse}{rgb}{0.00,0.50,0.50}

\definecolor{gesfblue}{rgb}{0.08,0.42,0.76}

\definecolor{gesfred}{rgb}{1,0,0}

\definecolor{gesfwhite}{rgb}{1,1,1}

\definecolor{gesfblack}{rgb}{0,0,0}

\newcommand{\geqn}[1]{\hypersetup{linkcolor=blue}(\ref{#1})\hypersetup{linkcolor=blue}}
\newcommand{\gfig}[1]{{\hypersetup{linkcolor=violet}Fig.~\ref{#1}\hypersetup{linkcolor=blue}}}

\graphicspath{{figs/}}

\begin{document}

\preprint{\hspace{-6mm} \today \hspace{6cm} IPMU18-0206, FERMILAB-PUB-18-487-T, [\href{https://arxiv.org/abs/1812.08376}{arXiv:1812.08376}]}

\title{The Scalar Non-Standard Interactions in Neutrino Oscillation}

\author{Shao-Feng Ge}
\affiliation{Kavli IPMU (WPI), UTIAS, University of Tokyo, Kashiwa, Chiba 277-8583, Japan}
\affiliation{Department of Physics, University of California, Berkeley, CA 94720, USA}
\affiliation{T.~D.~Lee Institute, Shanghai 200240, China}
\affiliation{School of Physics and Astronomy, Shanghai Jiao Tong University, Shanghai 200240, China}
\affiliation{Theoretical Physics Department, Fermi National Accelerator Laboratory, Batavia, IL 60510, USA}
\author{Stephen J. Parke}
\affiliation{Theoretical Physics Department, Fermi National Accelerator Laboratory, Batavia, IL 60510, USA}

\begin{abstract}
The scalar nonstandard interactions (NSI) can also introduce matter effect for neutrino
oscillation in a medium. Especially the
recent Borexino data prefers nonzero scalar NSI, $\eta_{ee} = - 0.16$.  In contrast to the
conventional vector NSI, the scalar
type contributes as a correction to the neutrino mass matrix rather than the matter potential.
Consequently, the scalar matter effect is energy independent while the
vector one scales linearly with neutrino energy.
This leads to significantly different phenomenological consequences in reactor,
solar, atmospheric, and accelerator neutrino oscillations. A synergy
of different types of experiments, especially those with matter density variation,
is necessary to identify the scalar NSI and guarantee the measurement of CP violation
at accelerator experiments.
\end{abstract}

\maketitle 

{\it Introduction} --
Neutrino oscillation was originally proposed to happen in vacuum due to
a nontrivial neutrino mass matrix \cite{nuOsc}. The matter potential
that neutrinos experience when propagating through matter medium was recognized
by Wolfenstein
in 1978 \cite{Wolfenstein:1977ue}. If the matter density times the neutrino energy is at the right
value, mixing angles can resonant to the maximal value and significantly change
the oscillation behavior \cite{resonant}.
The MSW effect \cite{Wolfenstein:1977ue, resonant} successfully
explains the observed solar neutrino fluxes \cite{solar} and makes a consistent
picture with the terrestrial experiments \cite{globalFit}.
Matter
effects have played a very important role in our understanding of neutrino
oscillations.

In the very first paper \cite{Wolfenstein:1977ue} on the neutrino matter effect,
Wolfenstein introduced non-standard interactions with generally
parametrized vector and axial-vector currents. This opens up the possibility
of using neutrino oscillation to probe extra new physics beyond the Standard
Model (SM), in addition to the neutrino mass matrix \cite{NSI-rev}.
The effect of NSI creates problems for the Dirac CP
phase measurement. Both real diagonal and complex off-diagonal elements can
fake the CP violation effect and hence disguise the genuine Dirac CP phase
\cite{Ge:2016dlx}.

Not only the ordinary matter can induce a matter effect, but also a dark sector
medium such as dark energy \cite{DE}, or MaVaNs (Mass Varying Neutrinos)
\cite{MaVaNs}, fuzzy dark matter (DM) \cite{fuzzyDM}, or with
a stand-alone particle \cite{standalone}. Although the dark sector density is much
lower than the ordinary matter density in the Sun or Earth, a large enough effect
is possible for super-light mediators. In the extreme environment of supernova,
NSIs can affect the collective oscillation of neutrinos \cite{SN}.

Coming back to the effect of NSI induced by ordinary matter on neutrino
oscillations, the discussion so far has been focusing on vector currents,
either from a vector mediator or with Fierz transformation from a charged scalar \cite{charged-scalar}.
Neutrinos can couple to not only vector field, but also scalar field.
To some extent, neutrino coupling with scalar field is even
more natural than the vector one. Because of the observed oscillation,
at least two of the three light neutrinos are massive. A natural mechanism for
neutrinos to acquire masses is via coupling with a scalar that has nonzero
vacuum expectation values. Such a possibility cannot be easily excluded.
Since the left-handed neutrino belongs to $SU(2)_L$ doublets and hence the scalar
coupling with neutrino may also couple with charged leptons, for example
via mixing with the neutral Higgs boson, it is natural to
see the matter effect induced by such a scalar particle.

We point out in this letter that the scalar NSI can introduce rich
phenomenology in reactor, solar, atmospheric, and accelerator
experiments. It is inevitable to use a synergy of multiple experiments to
test the scalar NSI and guarantee the CP measurement against the scalar NSI.
Although the matter effect can also arise from scalar mediator in MaVaNs \cite{MaVaNs-osc}
and fuzzy dark matter \cite{fuzzyDM} scenarios, its size is not proportional
to the ordinary matter density. Either the matter effect is modulated by dark energy,
neutrino, or dark matter densities, or is proportional to a nontrivial function
of the matter density such as $\tanh \rho$. In contrast,
we make model-independent study of the scalar NSI that scales 
with constant proportionality to the ordinary matter density and has rich
phenomenological consequences.

\vspace{2mm}
{\it Matter Potential and NSI} -- 
Before exploring the scalar NSI, let us first take a look how the vector
interactions can introduce matter potential in neutrino oscillation \cite{MatterEffect}.
In the SM, the matter potential can be induced by both charged and neutral currents.
With Fierz transformation \cite{Fierz}, the effective Lagrangian contributed by
charged current
$\propto
\left[
  \bar e(p_1) \gamma^\mu P_L \nu_e(p_2)
\right]
\left[
  \overline{\nu_e}(p_3) \gamma^\nu P_L e(p_4)
\right]
$
turns into the neutral current form \cite{Mohapatra:1998rq}
\begin{equation}
  \mathcal L^{\rm eff}_{\rm cc}
=
- \frac {4 G_F}{\sqrt 2}
\left[
  \overline{\nu_e}(p_3) \gamma_\mu P_L \nu_e(p_2)
\right]
\left[
  \overline e(p_1) \gamma^\mu P_L e(p_4)
\right].
\label{eq:Leff}
\end{equation}
Here we use electrons as the environmental matter for illustration
and the same procedure applies for protons and neutrons.
The matter effect comes from the so-called forward scattering
that experiences zero momentum transfer, $p_1 = p_4 \equiv p_e$ and
$p_2 = p_3 \equiv p_\nu$, and hence is momentum independent.
We can obtain the matter potential by
sandwiching the effective Lagrangian with in and out states,
\begin{subequations}
\begin{eqnarray}
  V_{cc}
& = &
- \langle \nu_e e(p_e, s_e) | \mathcal L^{\rm cc}_{\rm eff} | \nu_e e(p_e, s_e) \rangle \,,
\\
  \overline V_{cc}
& = &
- \langle \bar \nu_e e(p_e, s_e) | \mathcal L^{\rm cc}_{\rm eff} | \bar \nu_e e(p_e, s_e) \rangle \,,
\end{eqnarray}
\end{subequations}
for the neutrino and anti-neutrino cases, respectively. With the neutrino and
electron spinors separated, we can first focus on the neutrino parts,
$\langle \nu_e | \overline \nu_e \gamma_\mu P_L \nu_e | \nu_e \rangle$ and
$\langle \bar \nu_e | \overline \nu_e \gamma_\mu P_L \nu_e | \bar \nu_e \rangle$.
The quantized neutrino spinor is $\nu_e \equiv a u + b^\dagger v$ where
$a$ and $b$ are the annihilation operators of neutrino and anti-neutrino, respectively.
Correspondingly, the external states are
$|\nu_e \rangle \equiv a^\dagger |0\rangle$ and
$|\bar \nu_e \rangle \equiv b^\dagger |0\rangle$. The anti-communication property
of the neutrino operators leads to opposite sign between the neutrino and anti-neutrino
matter potentials $V_{\rm cc} = - \overline V_{\rm cc} = \sqrt 2 G_F n_e$.

For the vector NSI, the only difference is
introducing an extra vector mediator between various neutrino flavors. In all
these cases, the matter effect contributes to the $\bar \nu \gamma^0 \nu$ term,
\begin{equation}
  \bar \nu_\beta
\left[
  \left( i \partial_t \mp V^{(\dagger)}_{\alpha \beta} \right) \gamma^0
- i \triangledown \cdot \gamma
- M_{\beta \alpha}
\right] \nu_\alpha
=
  0 \,,
\label{eq:EOM}
\end{equation}
with $- V_{\alpha \beta}$ for neutrino and $V^\dagger_{\alpha \beta}$ for
anti-neutrino. When expanded to the linear order, the Hamiltonian
for neutrino oscillation is
\begin{equation}
  \mathcal H
\approx
  E_\nu
+ \frac {M M^\dagger}{2 E_\nu}
\pm (V_{\rm SI} + V_{\rm NSI}) \,,
\label{eq:H}
\end{equation}
where the neutrino mass matrix $M$, the matter potential $V_{\rm SI}$ for
the standard interaction (SI), and the NSI counterpart $V_{\rm NSI}$ are
$3 \times 3$ matrices. For simplicity, we denote $|p_\nu| \approx E_\nu$ since
neutrino masses are tiny.
Note that the neutrinos (anti-neutrinos) oscillation
is described by $\mathcal H$ ($\mathcal H^*$), respectively. The matter effect
appears as an extra term added to the Hamiltonian or energy.
Both the SI and NSI act as matter {\it potential}. For the matter potential to be
important, either the neutrino energy or the matter density
should be large enough, $2 E_\nu V \gtrsim \Delta m^2_{ij}$ where the latter has
two typical values, $\Delta m^2_{21} = 7.5 \times 10^{-5}\,\mbox{eV}^2$ or
$\Delta m^2_{31} = 2.7 \times 10^{-3}\,\mbox{eV}^2$.

\vspace{2mm}
{\it The Scalar NSI} -- The scalar NSI effect is no longer a matter potential.
This is because the effective Lagrangian \geqn{eq:Leff} is no longer composed of
vector current but a scalar Yukawa term for Dirac neutrinos
\begin{equation}
  \mathcal L^s_{\rm eff}
\propto
  y_f Y_{\alpha \beta}
\left[
  \bar \nu_\alpha(p_3) \nu_\beta(p_2)
\right]
\left[
  \overline f(p_1) f(p_4)
\right] \,,
\end{equation}
which cannot convert to vector currents \cite{Fierz-scalar}. In the nonrelativistic
limit the environmental fermion spinor reduces to $u_f = (\xi, \xi)^T$ where
$\xi_+ = (1, 0)^T$ or $\xi_- = (0, 1)^T$ for the two spin polarizations.
Consequently, $\bar u_f u_f = \bar u_f \gamma^0 u_f = 2 \xi^\dagger \xi = n_f$,
reducing to the matter density.
The correction to the Dirac equation \geqn{eq:EOM} then shifts to the
mass term
\begin{equation}
  \bar \nu_\beta
\left[
  i \partial_\mu \gamma^\mu
+
\left(
  M_{\beta \alpha}
+ \frac {\sum_f n_f y_f Y_{\alpha \beta}}{m^2_\phi}
\right)
\right] \nu_\alpha
=
  0 \,,
\end{equation}
where $y_f$ is the Yukawa coupling of the scalar mediator $\phi$ with
the environmental fermion $f$,
$Y_{\alpha \beta}$ is the one with neutrinos,
$Y_{\alpha \beta} \bar \nu_\alpha \phi \nu_\beta + h.c.$,
and for a real scalar $Y = Y^\dagger$.
Defining $\delta M \equiv \sum_f n_f y_f Y / m^2_\phi$, the effective
Hamiltonian \geqn{eq:H} becomes
\begin{equation}
  \mathcal H
\approx
  E_\nu
+ \frac {\left( M + \delta M \right) \left( M + \delta M \right)^\dagger}{2 E_\nu}
\pm V_{\rm SI} \,.
\label{eq:Hs}
\end{equation}
The scalar NSI appears as a correction to the neutrino mass term, rather than
the matter potential. Since matter effect is inversely proportional to the mediator mass,
$1 / m^2_\phi$, large enough $\delta M$ is possible with a light enough
scalar that can
survive current constraints from supernova \cite{SN-constraint}, meson
and lepton decays \cite{Pasquini:2015fjv},
neutrino electron scattering \cite{Sevda:2016otj},
neutrino trident production \cite{nuTrident},
Big Bang Nucleosynthesis \cite{BBN},
coherent scattering \cite{Coherent},
stellar cooling \cite{Stellar},
and other various observables \cite{Batell:2016ove}.
None of them can exclude the parameter region with both the mediator mass
and couplings being tiny.

The neutrino mass matrix $M = V_\nu D_\nu V^\dagger_\nu$ can be diagonalized by
the mixing matrix $V_\nu \equiv P_\nu U_\nu Q_\nu$ where $P_\nu$ and $Q_\nu$
are diagonal rephasing matrices, $U_\nu$ the PMNS matrix, and
$D_\nu \equiv \mbox{diag}\{m_1, m_2, m_3\}$ is the diagonal mass matrix.
With $\delta M$, the unphysical rephasing matrix $P_\nu$ can not be rotated away.
To comply with the established conventions,
we rotate $P_\nu$ into the scalar NSI contribution,
$M \rightarrow \widetilde M \equiv U_\nu D_\nu U^\dagger_\nu$
and $\delta M \rightarrow \delta \widetilde M \equiv P^\dagger_\nu \delta M P_\nu$.
For easy comparison with the genuine mass term, we parametrize the scalar NSI as
\begin{equation}
  \delta \widetilde M
\equiv
  \sqrt{\Delta m^2_{31}}
\left\lgroup
\begin{matrix}
  \eta_{ee}     & \eta_{e \mu}    & \eta_{e \tau}   \\
  \eta_{\mu e}  & \eta_{\mu \mu}  & \eta_{\mu \tau} \\
  \eta_{\tau e} & \eta_{\tau \mu} & \eta_{\tau \tau}
\end{matrix}
\right\rgroup \,,
\label{eq:dM}
\end{equation}
where $\eta_{\alpha \beta}$ are dimensionless parameters. Note that
those phases in $\delta \widetilde M$ are combinations of the unphysical
phases from $P_\nu$ and the scalar NSI matrix $\delta M$.
In the presence of the scalar NSI, the unphysical phases in $P_\nu$
can also have physical consequences on neutrino oscillation. 
The same thing happens to the absolute mass scale
that can be subtracted from $M M^\dagger$
as a common $m^2_i$ term but is always associated
with the scalar NSI contribution as $M \delta M^\dagger + M^\dagger \delta M$.
In this sense, the scalar NSI is also totally different from the vector NSI.

The vector NSI always conserves chirality which is no longer true for the
scalar NSI. The latter can only appear as correction
to the neutrino mass term that flips chirality \cite{SpinFlip,Liao:2015rma},
contrary to previous studies \cite{nuNSI-vec-scalar}.
Note that the scalar NSI is not suppressed by either the non-relativistic
environmental fermion $f$, $\langle m_f / E_f \rangle \approx 1$
\cite{SpinFlip}, or the relativistic neutrinos since neutrino oscillation
directly probes the neutrino mass term, $M + \delta M$.
As long as the correction $\delta M$ is comparable with the already
tiny neutrino mass term $M$, the scalar NSI can have a sizable effect.

\vspace{2mm}
{\it Phenomenological Consequences} -- 
Neutrino oscillation probes not only neutrino mixing but also
the neutrino interactions with the medium. Wolfenstein pointed out that
``{\it even if all neutrinos are massless it is possible to have
oscillations occur when neutrinos pass through matter}'' \cite{Wolfenstein:1977ue}.
He estimated ``{\it the oscillation length in matter of normal density
is of the order $10^9$\,cm}'' ($10^4$\,km), which is inversely proportional
to the matter potential. In the absence of the genuine mass term, the vector NSI
leads to an oscillation phase $e^{i V L}$ and hence the oscillation length 
$L \propto 1/V$ is independent of the neutrino energy but is only
a function of the medium density.

Nowadays, we have already measured neutrino oscillation attributed
to mass splittings that lead to an oscillation phase
$e^{i \Delta m^2_{ij} L / 4 E_\nu}$ and hence
the oscillation length is proportional to the neutrino energy,
$L \propto E_\nu / \Delta m^2_{ij}$. But the question is whether the measured
mass splittings are the genuine one $M$ from the fundamental
Lagrangian or the faked mass matrix $\delta M$ by the scalar NSI.
Even in the absence of the genuine mass matrix, oscillation
can still happen due to the scalar NSI \cite{Sawyer:1998ac}.
There is no essential difference between the genuine mass term and
the one induced by the scalar NSI.

\begin{figure}[t]
\centering
\includegraphics[height=4.2cm,width=3.5cm,angle=-90]{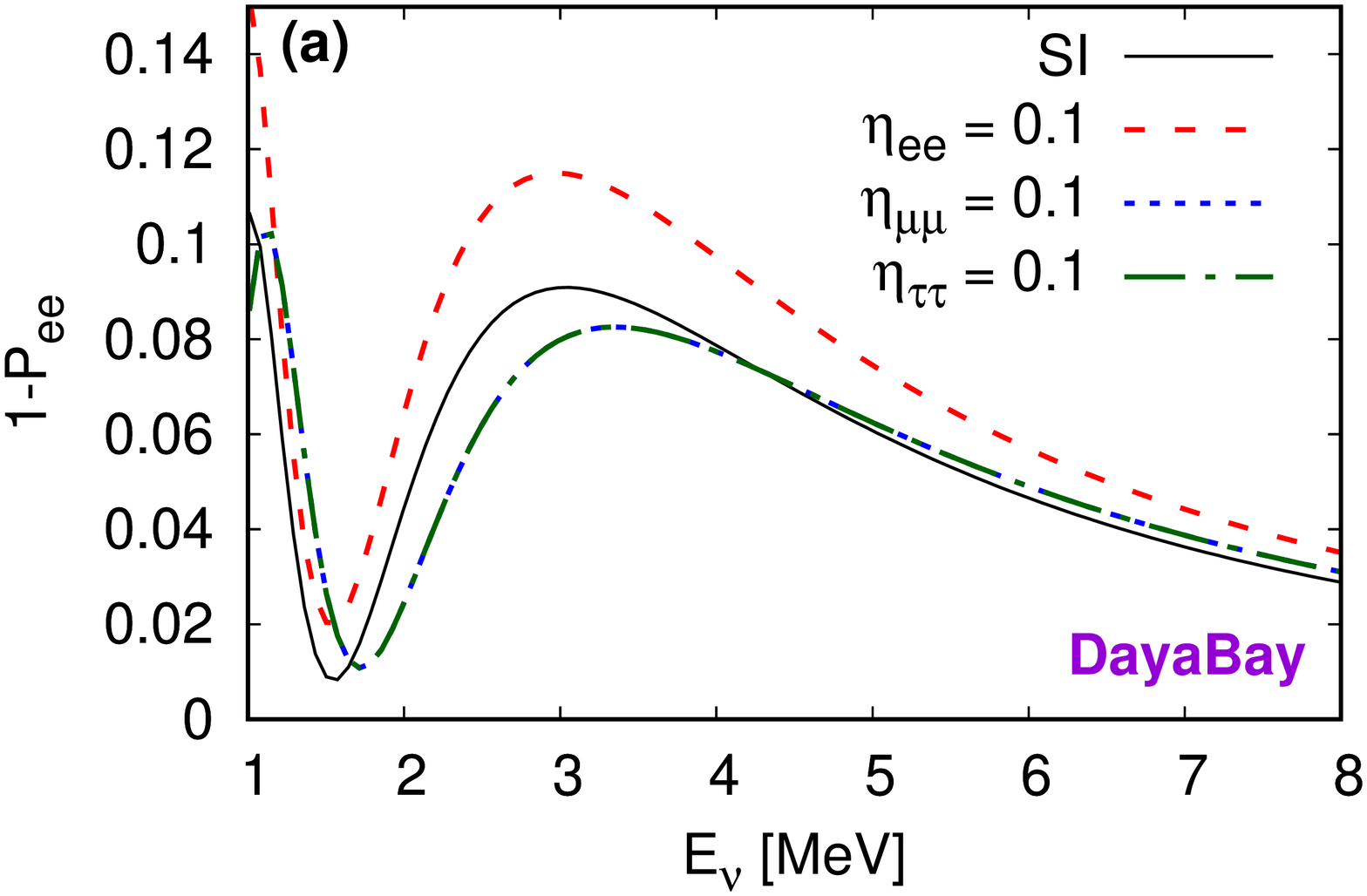}
\includegraphics[height=4.2cm,width=3.5cm,angle=-90]{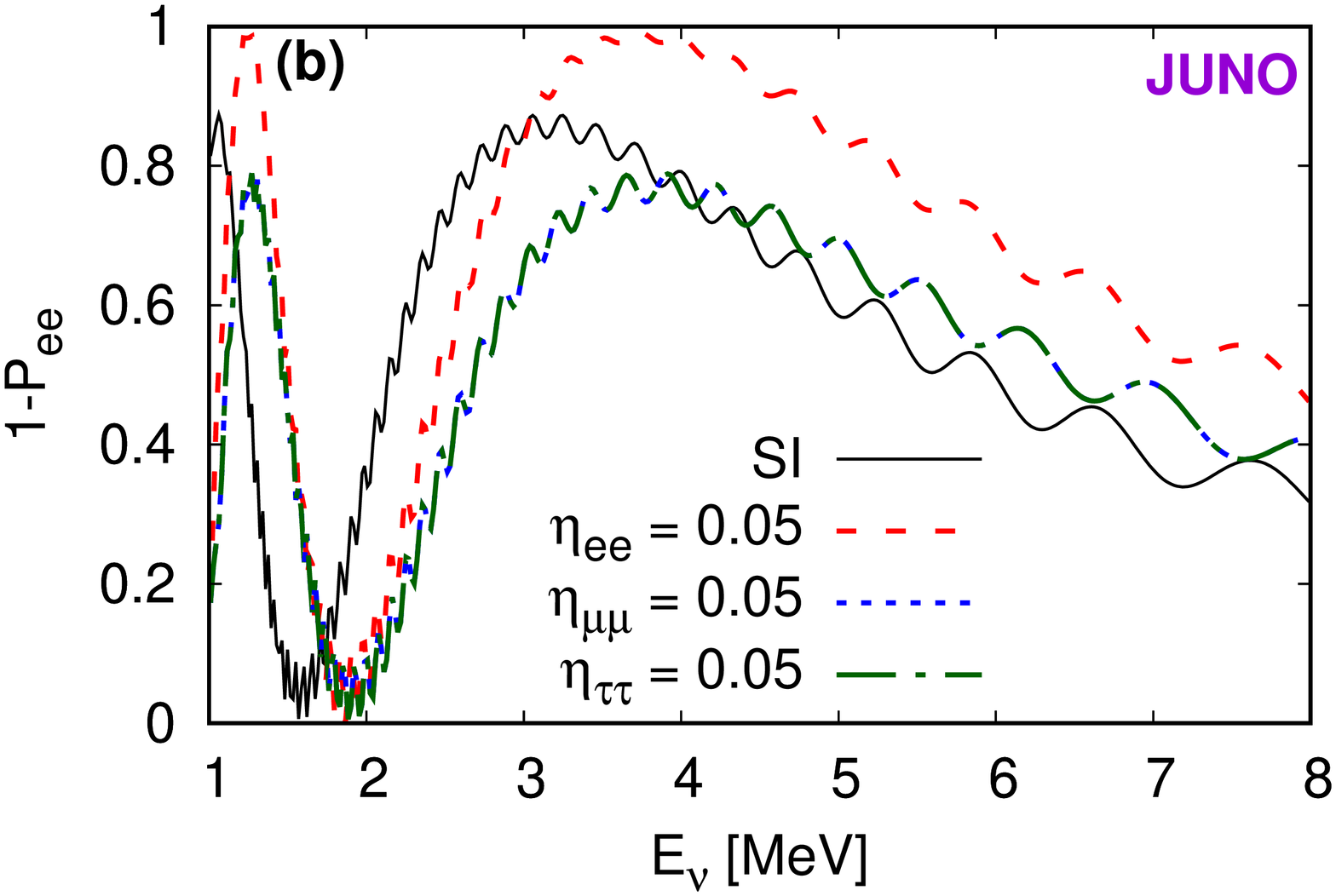}
\caption{The effect of the scalar NSI on the reactor anti-neutrino oscillation
         probabilities at (a) Daya Bay and (b) JUNO.}
\label{fig:dayabay}
\end{figure}

Its dependence on the matter density can help us to identify the scalar NSI.
While the genuine mass matrix $M$ is independent of environmental conditions,
the scalar NSI contribution $\delta \widetilde M$ scales with the matter density.
The oscillation probability can feel the matter density variations along the baseline.
For short-baseline terrestrial experiments, the variation
in the matter density is negligible and one combination of
$M$ and $\delta M$ can be redefined as the effectively measured mass matrix
$\widetilde M_{\rm re}$. Since the
rector experiments such as KamLAND \cite{KamLAND}, Daya Bay \cite{DayaBay},
and RENO \cite{RENO} give the most precise measurements, not to say the
future JUNO \cite{JUNO},
we implement the matter density subtraction at their typical matter density
$\rho_s = 2.6\,\mbox{g}/\mbox{cm}^3$ \cite{JUNO-matter},
\begin{equation}
  \widetilde M
+ \delta \widetilde M(\rho)
\equiv
  \widetilde M_{\rm re}
+ \delta \widetilde M(\rho_s) \frac {\rho - \rho_s}{\rho_s} \,,
\label{eq:subtraction}
\end{equation}
for both the neutrino and anti-neutrino modes. With $\rho = \rho_s$, the
effective mass matrix is exactly the reconstructed one
$\widetilde M_{\rm re} \equiv \widetilde M + \delta \widetilde M(\rho_s) = U_\nu D_\nu U^\dagger_\nu$
where $\delta \widetilde M(\rho_s)$ is actually \geqn{eq:dM} fixed at
the subtraction density $\rho_s$.
\begin{figure}[t]
\centering
\includegraphics[height=4.2cm,width=3.5cm,angle=-90]{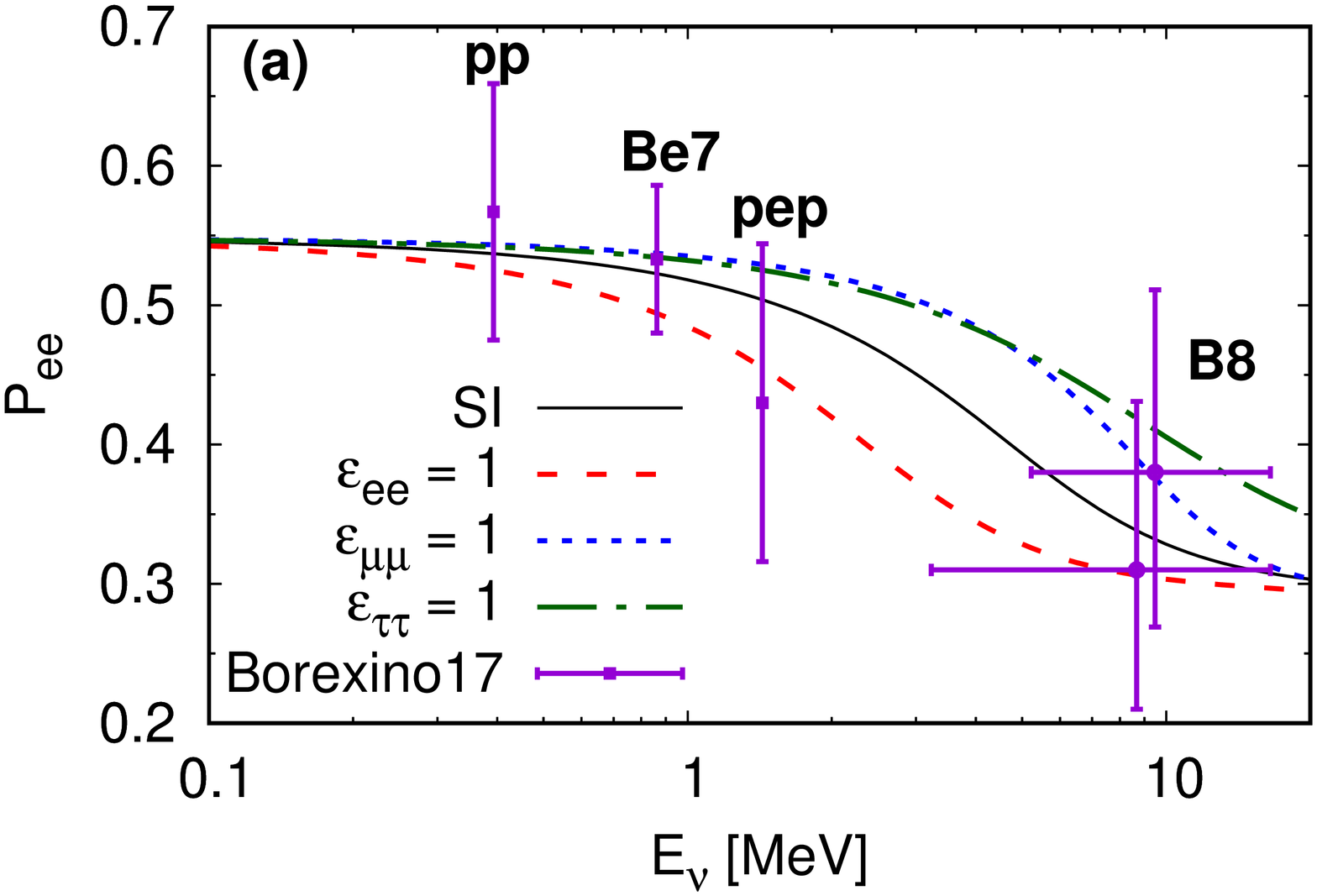}
\includegraphics[height=4.2cm,width=3.5cm,angle=-90]{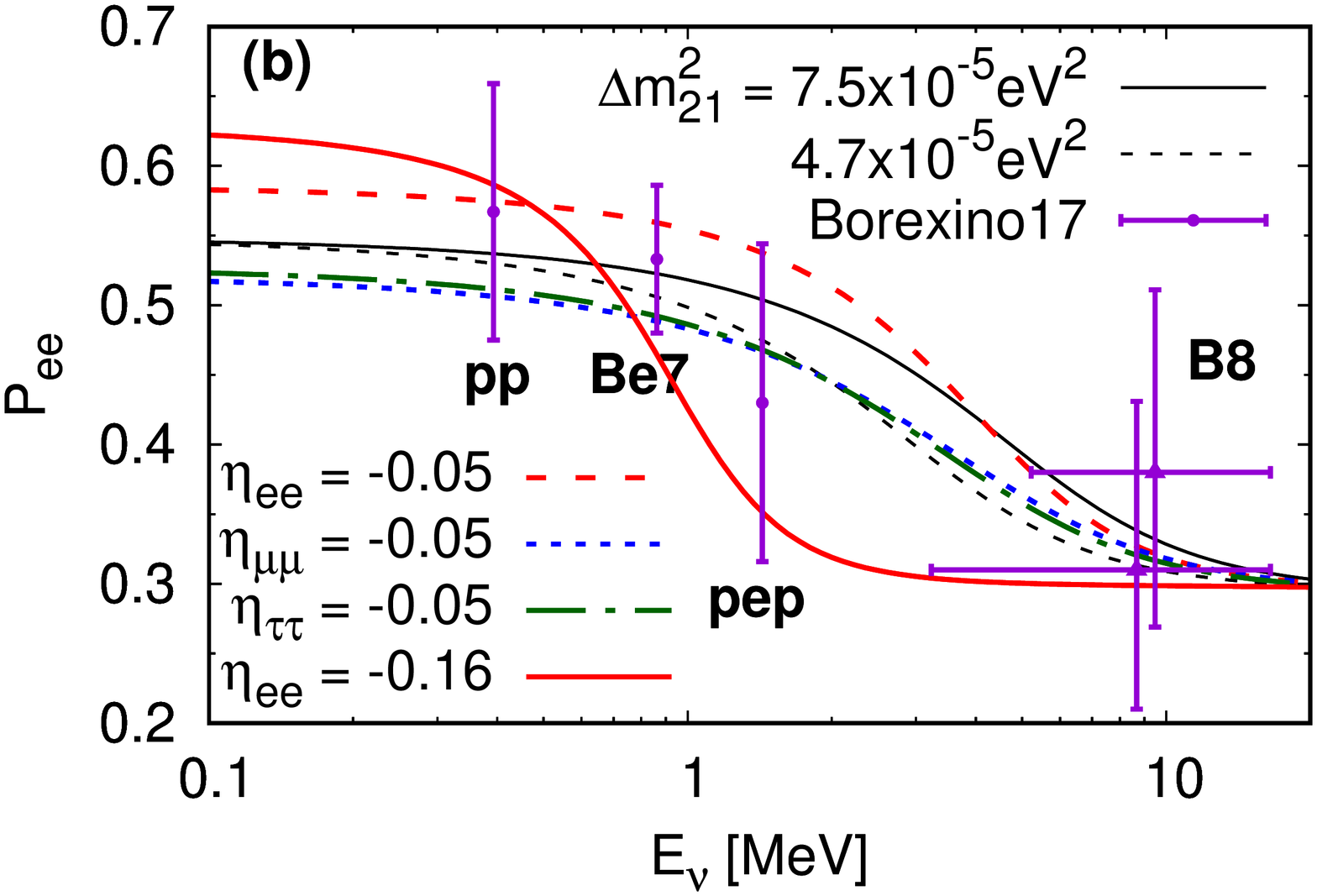}
\caption{The solar neutrino conversion probabilities with (a) the vector
         and (b) the scalar NSIs, together with the Borexino measurement \cite{Borexino17}
         of the $pp$, $^7$Be, and $pep$ fluxes.}
\label{fig:solar}
\end{figure}
We implement these in the NuPro package \cite{NuPro} for the simulations across
this paper. The \gfig{fig:dayabay} shows that the scalar NSI has large
effects at both Daya Bay and JUNO.
Nevertheless, the reactor experiments are not enough since the density subtraction
in \geqn{eq:subtraction} can eliminate the effects of the scalar NSI.
A global effort of combining different types of experiments is necessary
to distinguish the genuine neutrino mass matrix and the scalar NSI correction.
\begin{figure}[b]
\centering
\includegraphics[height=6cm,angle=-90]{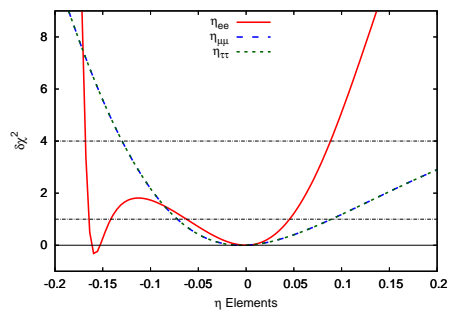}
\caption{The $\chi^2$ fit of the scalar NSI to the Borexino data \cite{Borexino17}.}
\label{fig:BorexinoFit}
\end{figure}

We show the effect of the scalar NSI
on the solar neutrino convertion in \gfig{fig:solar}, with a similar
plot of the vector NSI for comparison. Different from the vector one, the scalar
NSI is energy independent and hence not suppressed at low energy.
In addition, the SNO neutrinos experience much higher matter density \cite{Solar17}
than the KamLAND reactor neutrinos and hence can experience a sizable
second term in \geqn{eq:subtraction} with $\rho \gg \rho_s$. Both mixing
angles and mass squared differences are modified.
This provides a possibility of explaining the
discrepancy between the KamLAND \cite{KamLAND} and SNO \cite{SNO} measurements
of the neutrino mass splitting $\Delta m^2_{21}$. The blue ($\eta_{\mu \mu} = -0.05$)
and green ($\eta_{\tau \tau} = -0.05$) curves for the scalar NSI can mimic the black dashed
curve for the SI with the best fit value $\Delta m^2_{21} = 4.7 \times 10^{-5} \mbox{eV}^2$
\cite{solar-rev}. Last year, the Borexino experiment made
the first simultaneous measurement of the $pp$, $^7Be$, and $pep$ fluxes
\cite{Borexino17}. As a naive estimation, we plot the $\chi^2$ function
for these three data points in \gfig{fig:BorexinoFit}.
The $\eta_{\mu \mu}$ and $\eta_{\tau \tau}$ elements are effectively the
same for the solar neutrino oscillation probability and their best fit is the SM scenario. But for
the $\eta_{ee}$ element, there is a local minimal around $\eta_{ee} = -0.16$
beyond which the $\chi^2$ curve has a sharp increase due to the flip of mass
eigenvalues. The lastest Borexino data \cite{Borexino17} do favor a
nonzero scalar NSI. We show the oscillation probability curve with
$\eta_{ee} = -0.16$ in \gfig{fig:solar} for comparison. The preferred
nonzero $\eta_{ee}$ is mainly fixed by the smaller
central value of the $pep$ flux than the SI prediction.
Although the curve shape is quite different from the standard case, it is still
consistent with the Borexino data, including the $^8B$ flux from the earlier
measurement \cite{Bellini:2008mr}. The future SNO+ \cite{SNO+} and Jinping
\cite{JinPing} neutrino experiments are more precise and can help to pin it down.

\begin{figure}[t]
\centering
\includegraphics[height=0.23\textwidth,width=3.5cm,angle=-90]{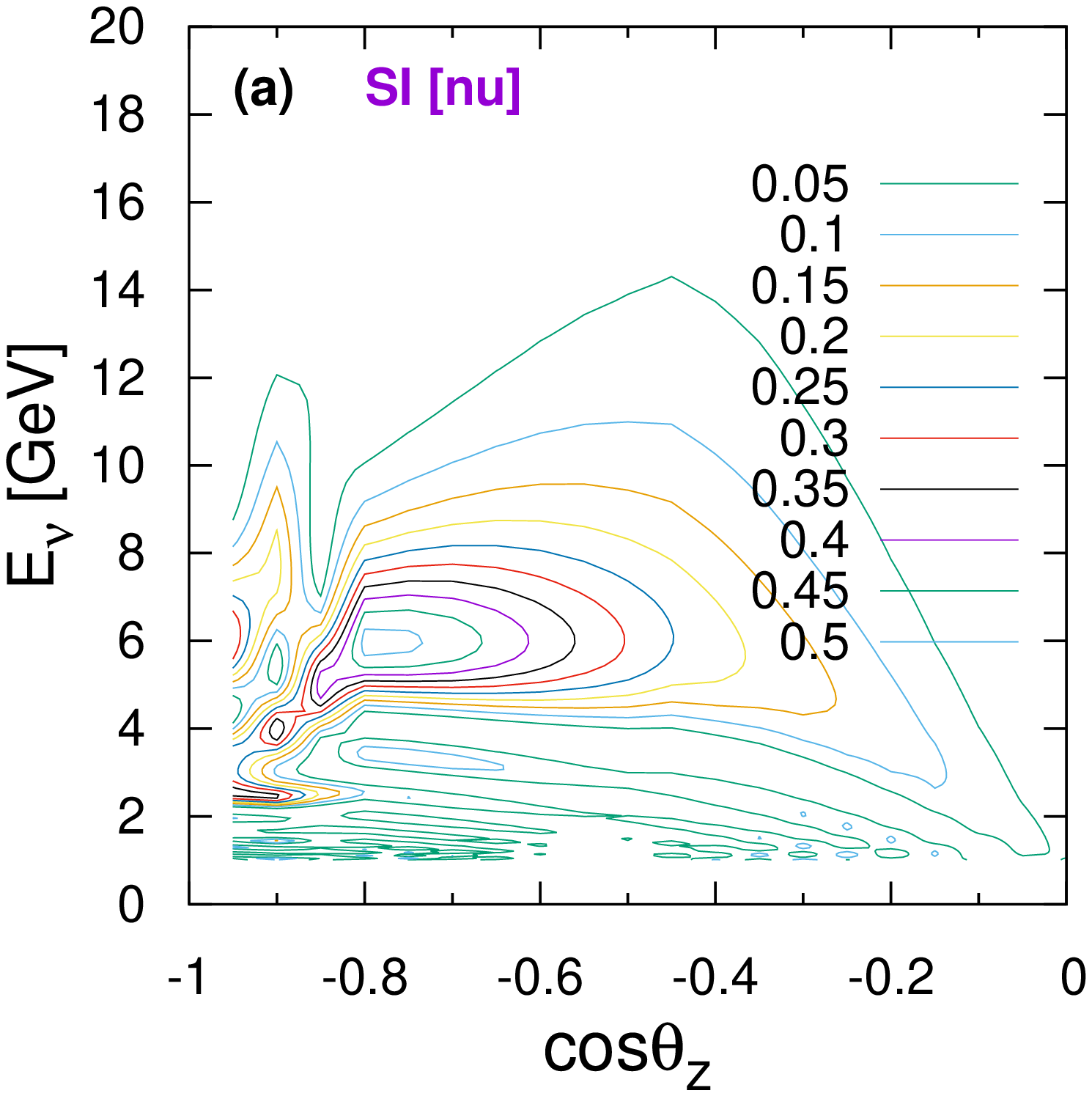}
\includegraphics[height=0.23\textwidth,width=3.5cm,angle=-90]{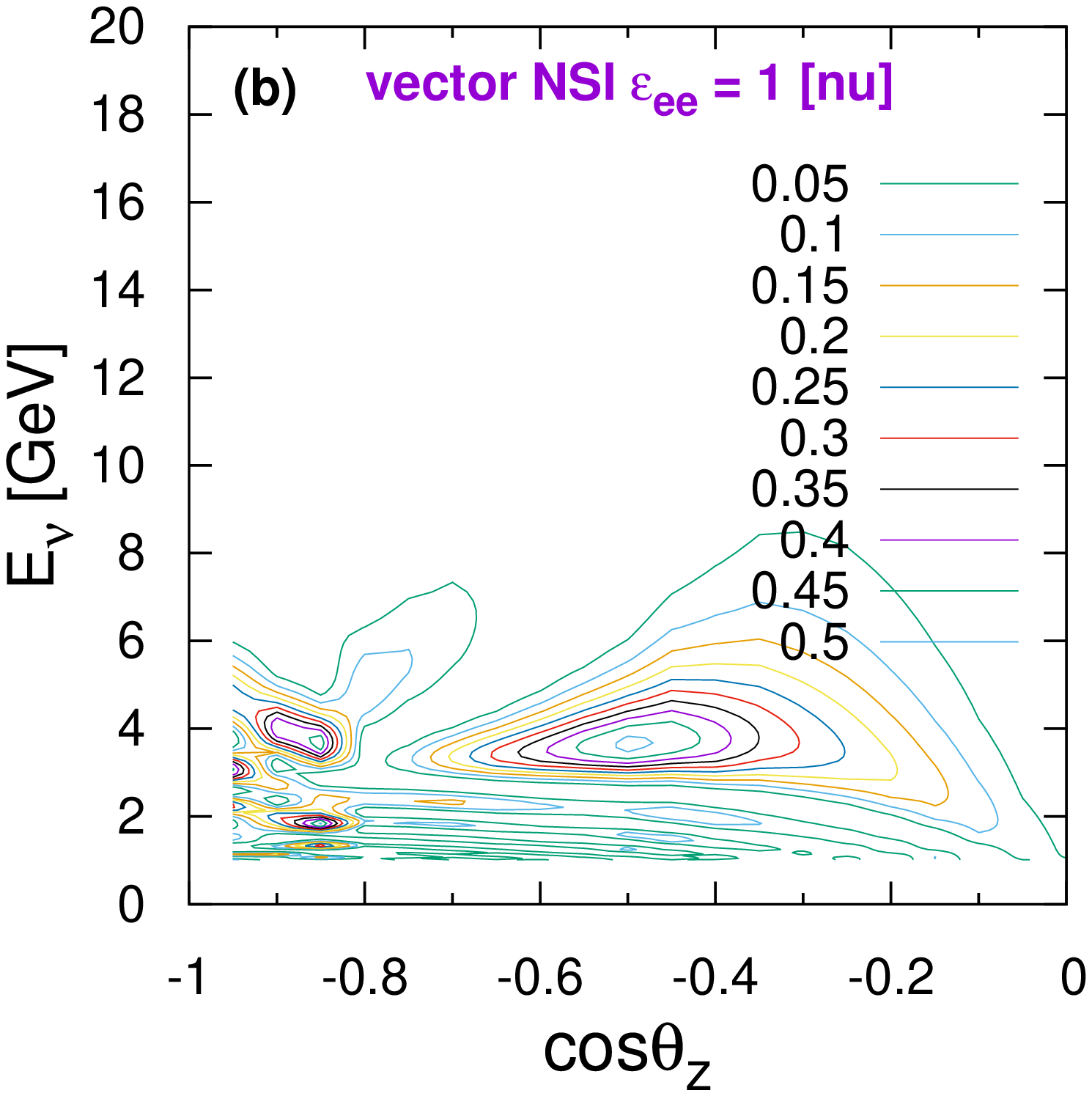}
\includegraphics[height=0.23\textwidth,width=3.5cm,angle=-90]{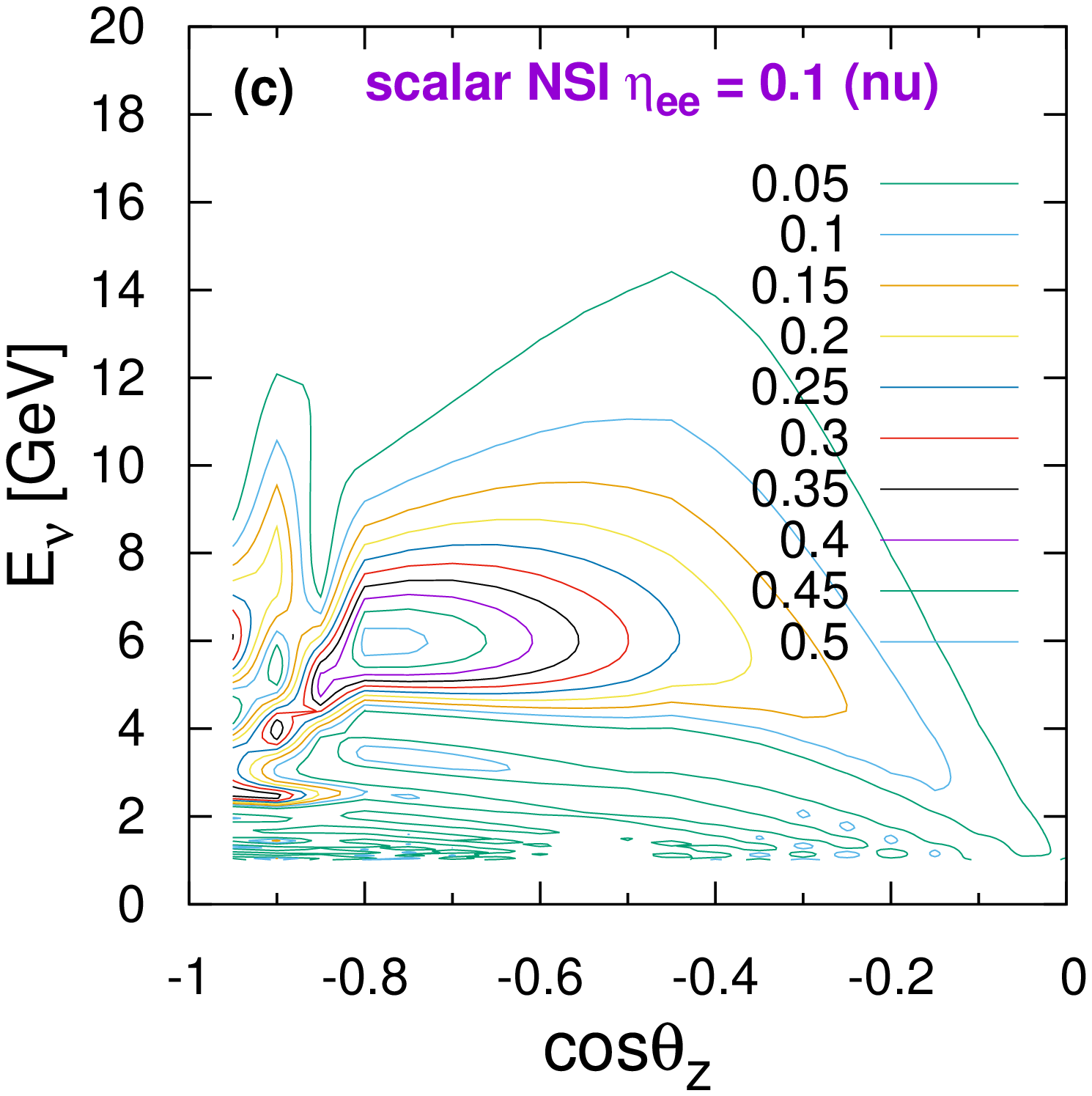}
\includegraphics[height=0.23\textwidth,width=3.5cm,angle=-90]{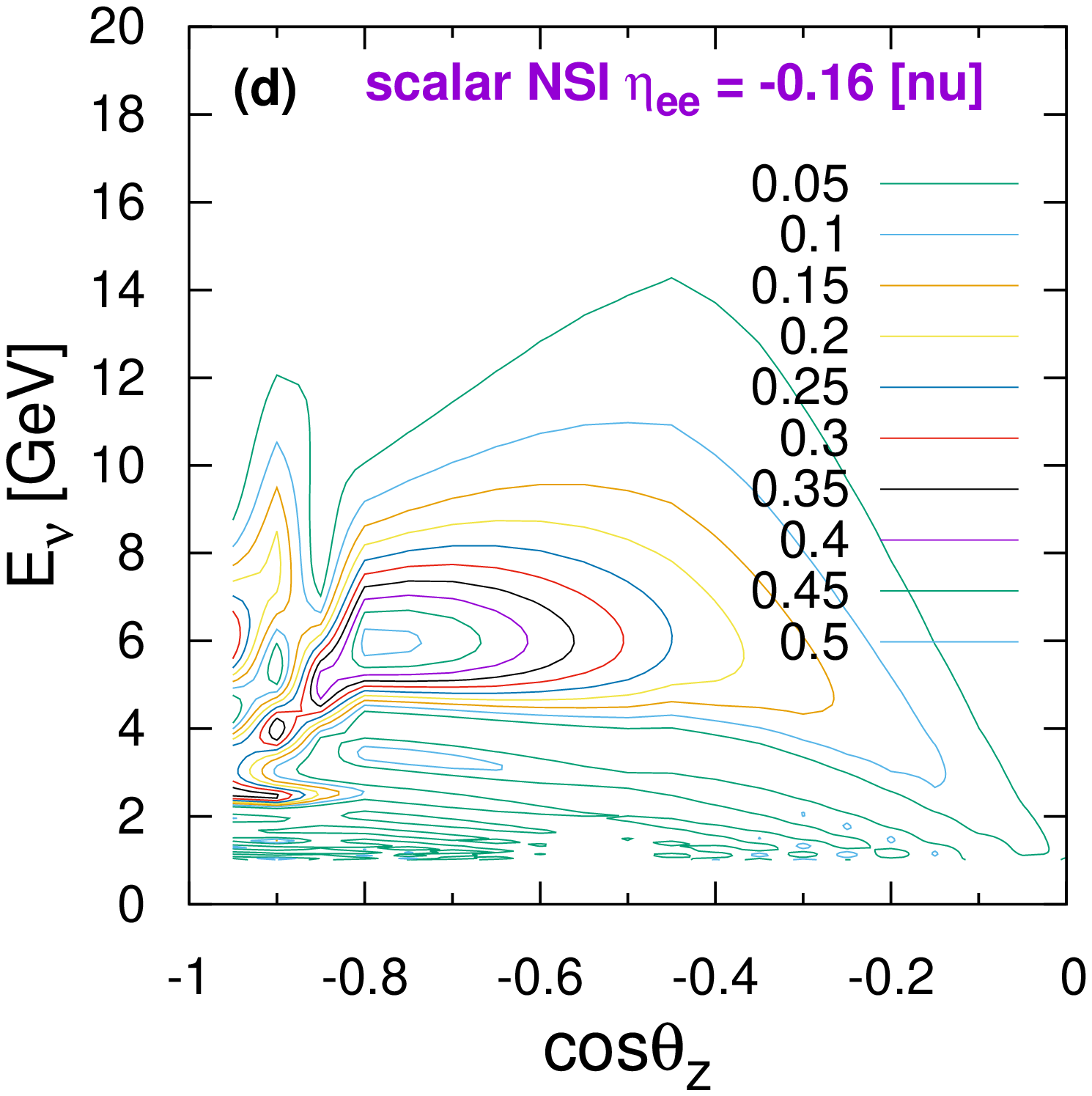}
\caption{The atmospheric neutrino oscillogram of $P_{\mu e}$ with (a) the SI only,
         (b) the vector NSI $\epsilon_{ee} = 1$, and the scalar NSI either
         (c) $\eta_{ee} = 0.1$ or (d) $\eta_{ee} = -0.16$ .}
\label{fig:atmos}
\end{figure}
The atmospheric neutrino oscillation can also experience matter density variation
and hence help identify the scalar NSI. In \gfig{fig:atmos} we show the
atmospheric neutrino oscillogram and its modification by the vector or scalar NSI.
Those neutrinos crossing the Earth core experience the most significant matter
density variation. Consequently, the core region ($\cos \theta_z \lesssim -0.8$)
shows the largest effect where the difference in $P_{\mu e}$ is as large
as $0.14$ between the SI and the scalar NSI. Note that the maximal value of $P_{\mu e}$
is around $0.5$ which is clearly seen in the decomposition formalism
\cite{decomposition}. The relative change is as large as $1$ in the energy
range $E_\nu \lesssim 5\,\mbox{GeV}$. We expect the PINGU
\cite{PINGU}, ORCA \cite{ORCA}, and INO \cite{INO} experiments
to put some constraints on the scalar NSI.
In addition, the lower energy threshold with Super-PINGU \cite{SuperPINGU}
can further enhance the sensitivity to the scalar NSI whose effect can
surpass the matter potential at lower energy.

\begin{figure}[t]
\centering
\includegraphics[height=4.2cm,width=3.5cm,angle=-90]{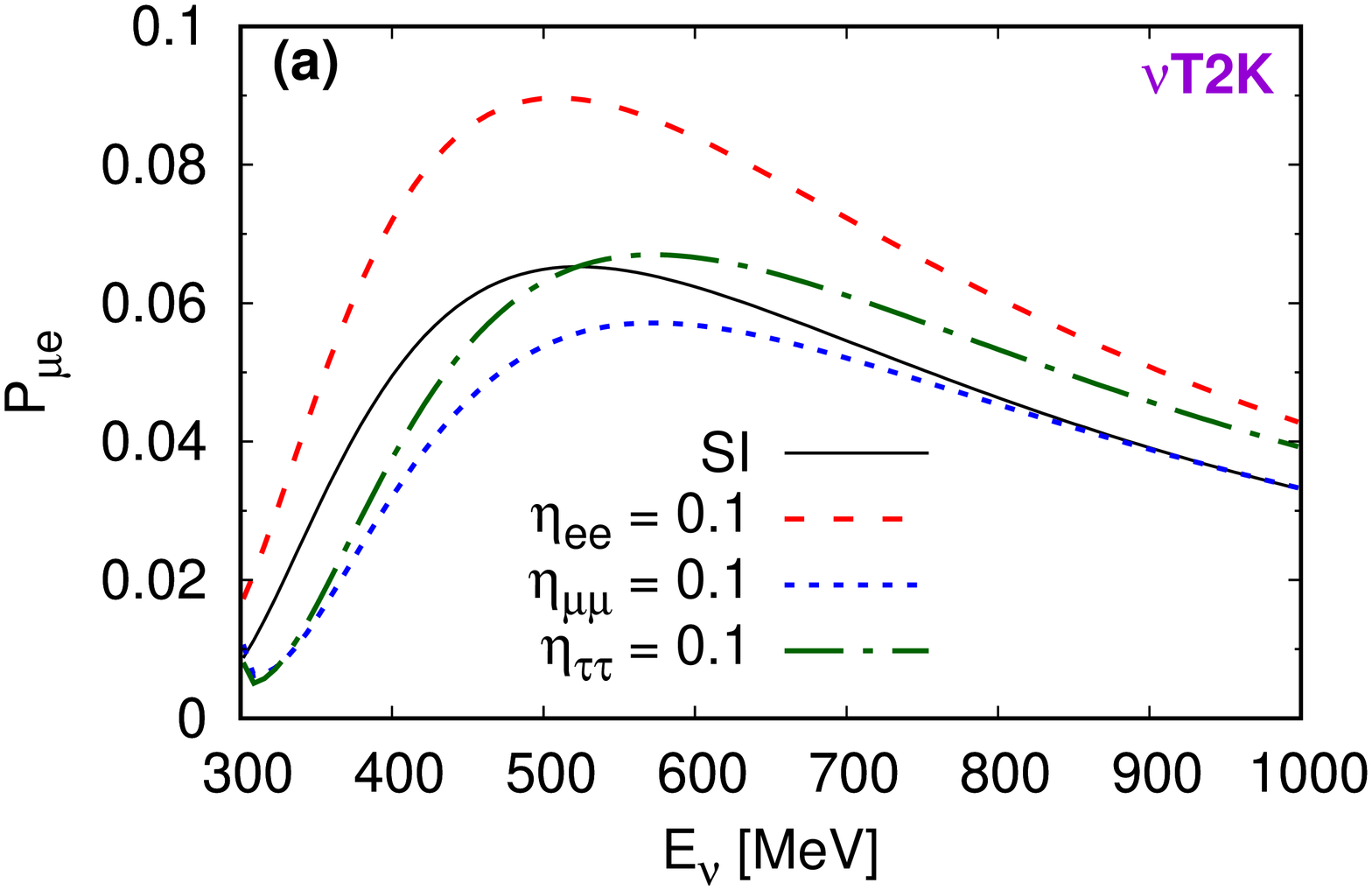}
\includegraphics[height=4.2cm,width=3.5cm,angle=-90]{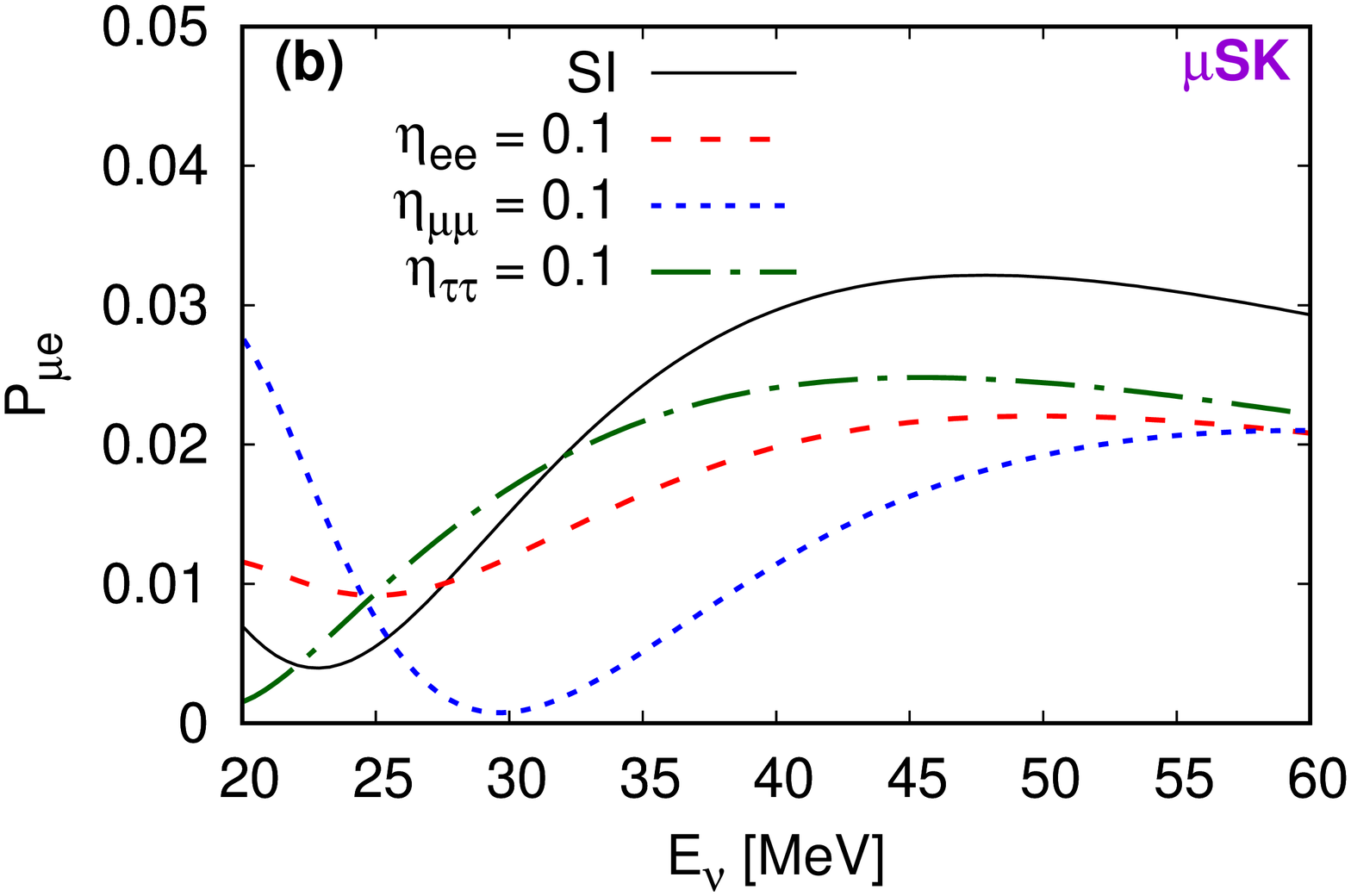}
\includegraphics[height=4.2cm,width=3.5cm,angle=-90]{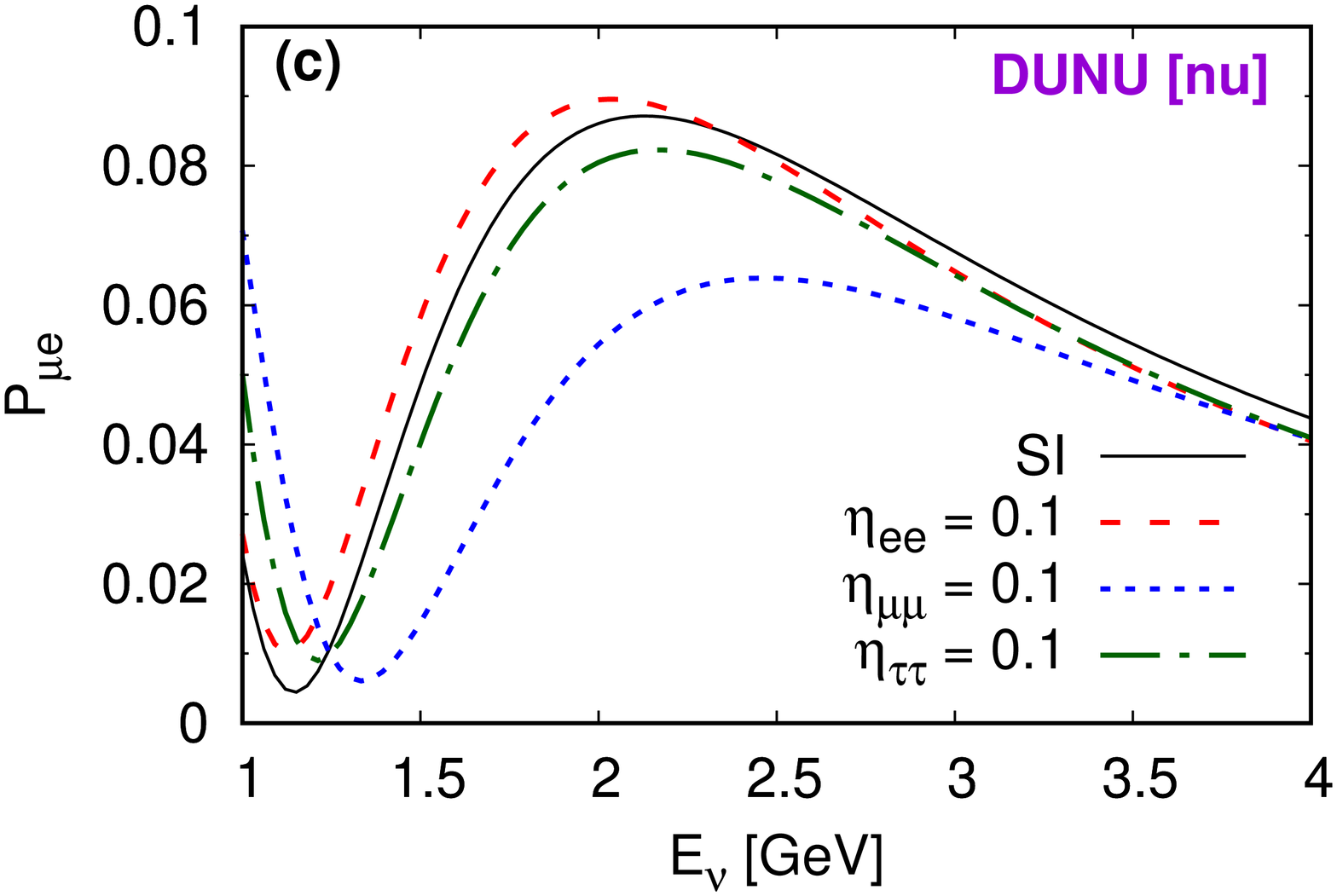}
\includegraphics[height=4.2cm,width=3.5cm,angle=-90]{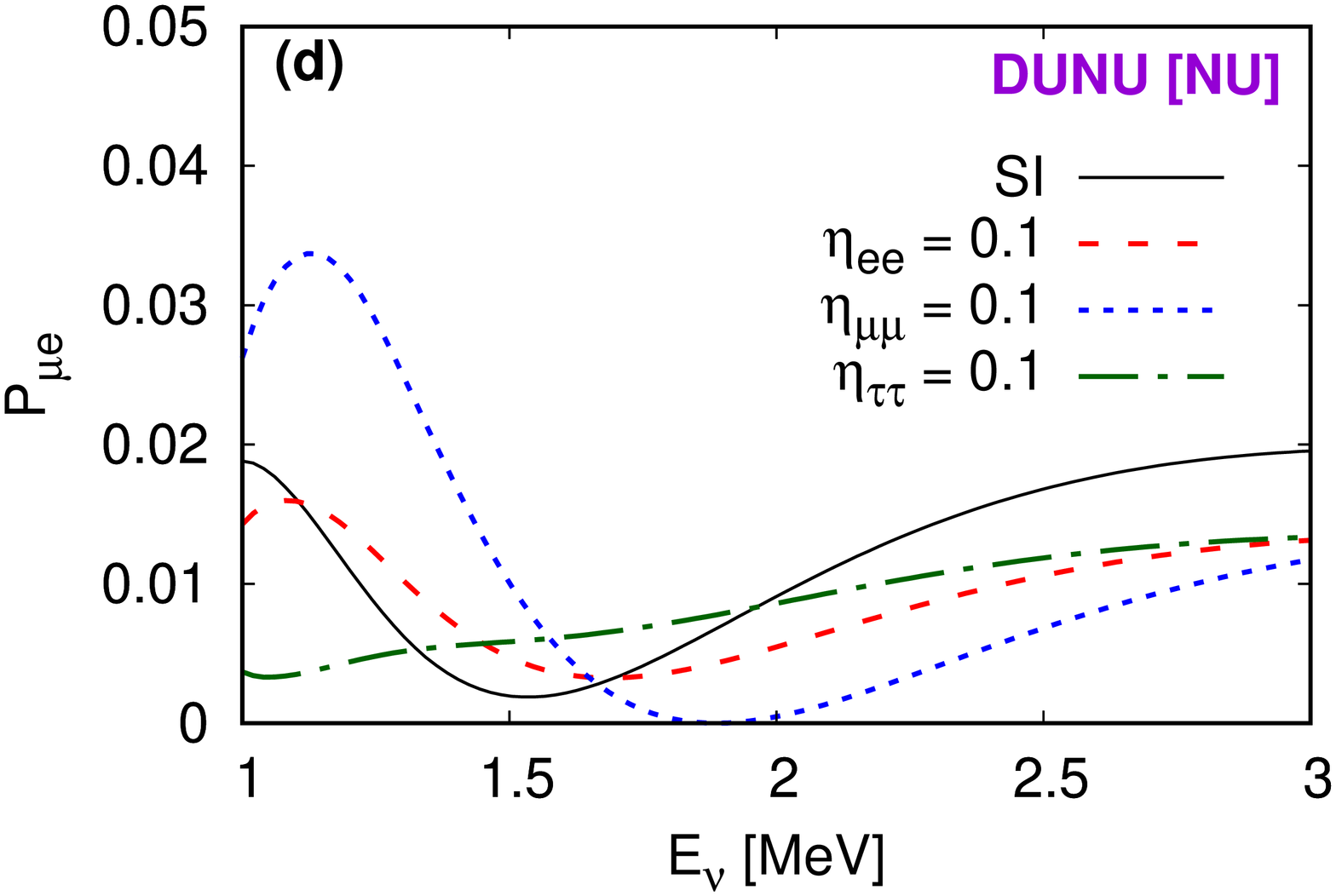}
\caption{The probability $P_{\mu e}$ at (a) $\nu$T2K, (b) $\mu$SK, as well as
         (c) neutrino and (d) anti-neutrino modes at DUNE.}
\label{fig:accelerator}
\end{figure}

For the accelerator neutrino experiments, whose main purpose is for the Dirac CP
phase measurement, the situation is a little more intricate. Since the effective
mass matrix is modified by the scalar NSI, the effective Dirac CP phase can be
quite different from the genuine one. We show the oscillation probability
$P_{\mu e}$ at TNT2K \cite{TNT2K},
including the neutrino mode $\nu$T2K and the anti-neutrino mode $\mu$SK,
and both modes at DUNE \cite{DUNE}
in \gfig{fig:accelerator}. To make the faked CP effect explicit, we remove
the matter density subtraction so that the effect of the scalar NSI appears in both
neutrino and anti-neutrino modes. Although the neutrino energy
varies a lot, all experiments receive comparable modification from the
scalar NSI which is totally different from the vector case.
If the complex phases and off-diagonal elements $\eta_{\alpha \beta}$ with
$\alpha \neq \beta$ are also introduced, the situation would become even more
complicated. The degeneracy between the genuine Dirac CP phase and
the scalar NSI can make the CP measurement more difficult, in the same
way as the vector NSI \cite{Ge:2016dlx} and the non-unitary mixing \cite{NUM}.

\vspace{2mm}
{\it Summary and Discussions} -- 
We point out that the scalar NSI has totally different features and
phenomenological consequences from the vector one.
The scalar NSI contributes as correction to the neutrino mass matrix
and hence its effect is independent of the neutrino energy.
Even for the low energy reactor, solar, and muon decay at rest
($\mu$DAR) experiments, the effect of the scalar
NSI cannot be ignored. In addition, the scalar NSI can fake the CP effect and
becomes a trouble to the on-going and future CP measurements at T2(H)K \cite{T2K},
NO$\nu$A \cite{NOvA}, and DUNE \cite{DUNE}. A global effort of using
matter-density-varying oscillation, such as the solar and atmospheric
neutrino experiments, is then needed for precision measurement of the
leptonic Dirac CP phase.

\section*{Acknowledgements}

We would like to thank Peter Denton, Jiajun Liao, and Alexei Smirnov
for useful discussions and references after reading an earlier version of this manuscript.

The work of SFG was supported by JSPS KAKENHI Grant Number JP18K13536,
World Premier International (WPI) Research Center Initiative, MEXT, Japan,
and Fermilab Neutrino Physics Center Fellowship Award Program. SFG is also
grateful to the hospitality of Patrick Huber and the Center for Neutrino
Physics at Virginia Tech where this paper was partially drafted.

This manuscript has been authored (SJP) by Fermi Research Alliance, LLC under Contract No.~DE-AC02-07CH11359 with the U.S. Department of Energy, Office of Science, Office of High Energy Physics.

This project (SJP) has received funding/support from the European Union's Horizon 2020 research and innovation programme under the Marie Sklodowska-Curie grant agreement No 690575 \& No 674896.

\end{document}